# Electrically driven strain-induced deterministic single-photon emitters in a van der Waals heterostructure

Jae-Pil So[1], Ha-Reem Kim[1], Hyeonjun Baek[1,2], Hoo-Cheol Lee[1], Woong Huh[3], Yoon Seok Kim[3], Kenji Watanabe[4], Takashi Taniguchi[5], Jungkil Kim[6], Chul-Ho Lee[3], and Hong-Gyu Park[1,3*]

[1]Department of Physics, Korea University, Seoul 02841, Republic of Korea.

[2]Center for Quantum Information, Korea Institute of Science and Technology, Seoul 02841, Republic of Korea.

[3]KU-KIST Graduate School of Converging Science and Technology, Korea University, Seoul 02841, Republic of Korea.

[4]Research Center for Functional Materials, National Institute for Materials Science, 1-1 Namiki, Tsukuba, Ibaraki 305-0044, Japan.

[5]International Center for Materials Nanoarchitectonics, National Institute for Materials Science, 1-1 Namiki, Tsukuba, Ibaraki 305-0044, Japan.

[6]Department of Physics, Jeju National University, Jeju 63243, Republic of Korea.

[*]Corresponding author. E-mail: hgpark@korea.ac.kr

**Abstract**

**Quantum confinement in atomically-thin TMDCs enables the realization of deterministic single-photon emitters. The position and polarization control of single photons have been achieved via local strain engineering using nanostructures. However, most existing TMDC-**

**based emitters are operated by optical pumping, while the emission sites in electrically pumped emitters are uncontrolled. Here, we demonstrate electrically driven single-photon emitters located at the positions where strains are induced by atomic-force-microscope indentation on a van der Waals heterostructure consisting of graphene, hexagonal-boron nitride, and tungsten diselenide. The optical, electrical, and mechanical properties induced by the local strain gradient were systematically analyzed. In particular, single-photon emission was observed at the indentation sites at 4 K. The emission exhibits photon anti-bunching behavior with a $g^{(2)}(0)$ value of ~0.3, intensity saturation and a linearly cross-polarized doublet. This robust spatial control of electrically driven single-photon emitters will pave the way for the practical implementation of integrated quantum light sources.**

**Introduction**

Quantum emitters created by natural or artificial defects in two-dimensional (2D) transition dichalcogenides (TMDCs) have been widely studied because of their unique 2D nature[1-3]. For example, the weak van der Waals forces between their layers enable TMDCs to be easily exfoliated and integrated with other structures, thereby allowing the development of the complex and functional quantum systems including single-photon devices[4,5]. TMDC-based emitters exhibit a very high light extraction efficiency compared to one-dimensional or three-dimensional emitters because of the absence of total internal reflection in the 2D geometry[6]. Furthermore, the electrical and optical properties of TMDCs can be modulated to large extents via strain engineering, owing to their small stiffness for out-of-plane displacements and their high strain limits[7-15]. Local tensile strains are efficiently induced in the TMDCs transferred onto nanostructures, resulting in exciton funneling and localized emission[16]. Thus, position-controlled optically-pumped single-photon emitter arrays have been successfully demonstrated using various nanostructures, including dielectric pillars[7,8], nanobubbles[9], optical waveguides[10-13], metal nanogaps[14], and metal nanoparticles[15]. More recently, deterministic control of the polarization in single-photon emission was achieved in a TMDC placed on a nanogap array[17]. The polarization direction changed depending on whether the nanogap was larger or smaller than a critical value.

The physical properties of TMDC-based emitters can be readily and reliably manipulated by electrical control[18-20]. Single-photon emitters in a TMDC heterostructure containing insulating and conducting layers such as h-BN and graphite can reflect the complexity and functionality of the potential landscape, which can be systematically controlled by band-structure engineering through the application of an electric field to the 2D heterostructure[21-23]. In particular, the demonstration of electrically driven single-photon devices based on tunnel junctions in graphene/h-BN/TMDC multilayers proves the feasibility of practical integrated quantum light sources[24-26]. However, despite the necessity of accurately positioning the emitters for next-generation on-chip quantum emitter devices, such positioning has not yet been achieved for electrically pumped emitters.

In this work, we demonstrate an electrically driven single-photon emitter array with deterministically controlled emission sites. A tunneling device based on a graphene/h-BN/WSe$_2$ heterostructure was placed on a deformable substrate, and a local tensile strain was applied on the 2D heterostructure by indentation using an atomic-force-microscope (AFM) tip[27]. A numerical simulation was performed to elucidate the indentation process. In the experiment, narrow electroluminescence (EL) emission peaks were obtained from the position-controlled emitters, which exhibited the single-photon features of photon anti-bunching with a g$^{(2)}$ of ~0.3, intensity saturation, and linearly cross-polarized zero-field splitting. We believe these electrically driven strain-induced single-photon emitters will be useful for the practical implementation of quantum technologies such as quantum communication and quantum information.

**Results**

**Design and fabrication of electrically driven single-photon emitters.** Figure 1A shows a schematic illustration of an electrically driven single-photon emitter in a vertical van der Waals heterostructure. A monolayer bottom graphene, few-layer hexagonal boron nitride (h-BN), monolayer WSe$_2$, and monolayer top graphene layers are stacked in turn from the bottom to the top on a deformable polymer substrate. Tensile strain is induced at the sites via indentation by an AFM tip[27]. Current is injected into the WSe$_2$ monolayer, which has a direct band gap, by applying a bias voltage between the top and bottom graphene layers.

The van der Waals heterostructure is rationally designed based on the following specifications. First, the bottom graphene is separated from the WSe$_2$ monolayer by the h-BN insulating layer; thus, the current can be injected uniformly into the entire area of the WSe$_2$ by tunneling. This current injection scheme is distinct from that in lateral split-gate p-n junction devices[28-30] and transient devices[31], in which EL emission was observed only near the source or gate contact edges. Second, an h-BN layer is not introduced between the top graphene and WSe$_2$ monolayer, because tunneling and current spreading are already achieved by the h-BN layer underneath the WSe$_2$[24]. In addition, the increase in the thickness of the heterostructure due to the introduction of an additional h-BN layer would have hindered the indentation of the WSe$_2$ monolayer. Third, the deformable substrate is useful for forming a permanent indentation site and preventing a return to the strain-free geometry. If an indentation site is fabricated in a van der Waals heterostructure on an inelastic substrate using the tip-based approach, it is inevitable that defects such as scratches and piercings in the structure will be formed[32].

The operating mechanism of the proposed device is schematically illustrated in Figure 1B. At zero bias between the bottom and top graphene layers, no charge flows through the vertical layers owing to the constant Fermi energy ($E_F$) across the heterojunction. When a bias voltage is applied, the $E_F$ increases above the conduction band edge of the monolayer $WSe_2$ and electron tunneling occurs through the h-BN barrier. This initiates the radiative recombination of electron-hole pairs residing in the optically active $WSe_2$ region. In particular, because the band gap is engineered by the induced strain, the electron-hole pairs funnel into the localized exciton trap and EL emission occurs at the indentation site.

Van der Waals heterostructures with indentation sites were fabricated via the following procedures (Fig. 1C). Exfoliated monolayer $WSe_2$ and few-layer h-BN were first placed in turn on monolayer graphene on a $SiO_2$/Si substrate using the conventional pick-and-place method with an adhesion material[33]. Another graphene monolayer was placed on top to cover the entirety of the monolayer $WSe_2$ (see Methods and fig. S1). We note that this heterostructure was stacked in the reversed order. Next, a polymer (PMMA) layer was spin-coated on the whole structure, and the structure was peeled off from the $SiO_2$/Si substrate via solution-based transfer method (Fig. 1C, (i)). The PMMA-coated heterostructure was turned upside-down as it floated on DI water. The inverted heterostructure on the PMMA layer was transferred and aligned on another $SiO_2$/Si substrate (Fig. 1C, (ii)). The sample was heated at 120 °C to anneal the bottom PMMA layer until it adhered to the target substrate. Finally, we formed indentation sites on the heterostructure using an AFM tip (Fig. 1C, (iii)). The adhesive interaction between the PMMA layer and the heterostructure led to the permanent elastic deformation of the indentation sites. The optical microscope image of the fabricated sample shows that all the layers were stacked as desired (Fig. 1D). We note that the indentation sites were created in the WSe$_2$ monolayer on the bottom graphene layer because the separation of the top and bottom graphene layers leads to efficient current injection into the region (Fig. 1E).

**Mechanical properties of the indented heterostructure.** The close-up AFM topography images of the indentation sites show the height profiles across the horizontal and vertical cross-sections. The heterostructure was permanently deformed with a depth of ~25 nm, which is the average value of the seven indentation sites fabricated by the AFM tip at the applied load of 6.33 μN (Fig. 2A). The maximum standard deviation of the depth was 6.9 nm, which was attributed to the heterogeneities in the WSe$_2$/h-BN/graphene/PMMA interfaces. The relatively large stiffness of the heterostructure, compared with that of the monolayer TMDC, led to the asymmetric shape of the indentation site: the tip could slide slightly along the surface instead of remaining at the same location during the indentation process. Nonetheless, this fabrication method effectively reduced the occurrence of uncontrolled wrinkles and random strain gradients, which were frequently observed in 2D layers transferred onto nanopatterned substrates, because only a small area of the AFM tip was in contact with the heterostructure[27]. In addition, the PMMA substrate held the heterostructure in place and forced it to follow the deformation contour, resulting in a highly localized strain field.

To further understand the indentation process, we performed a finite-element method (FEM) simulation to calculate the morphology of the indent on the heterostructure when it was pressed by an incompressible conical tip with a radius of curvature $r_t$ (see Methods). The van der Waals heterostructure was modeled as a single linear elastic material with effective mechanical parameters[34], and placed on the PMMA substrate (see Methods). The simulation shows that a conic-shaped deformation occurred at the indentation site (Fig. 2B). The indentation depth was 25 nm when the force applied by a tip with $r_t$ = 5 nm was 6,000 nN, which agrees well with the experimental result shown in Fig. 2A. Figure 2C shows the calculated indentation depth as a function of the force applied on the heterostructure on the PMMA substrate. The indentation depth increased and became saturated as the applied force increased. As the indentation deepened, the contact stiffness was dominated by the behavior of the polymer substrate. Because of the large strain limit for vertical displacements[35,36], plastic deformation of the PMMA substrate occurred, resulting in a saturation of the indentation depth. In addition, we calculated the line profiles of the diagonal strain component along the vertical direction ($|\varepsilon_{zz}|$) at various tip radii ($r_t$ = 5, 10 and 15 nm) at the fixed force of 6,000 nN (Fig. 2D). The calculated $|\varepsilon_{zz}|$ exhibited a higher and narrower peak when a sharper tip was used. Therefore, the shape and size of the tip can provide additional degrees of freedom in strain engineering[27].

**Measurement of single-photon emission.** To investigate the emission properties of the fabricated heterostructure with indentation sites, a high-resolution confocal raster scan was performed at 4 K using two scanning galvo mirrors, while a bias voltage was applied to the heterostructure. The EL emission was collected by a 100× objective lens with the numerical aperture of 0.85 and sent to avalanche photodiodes (APD) via optical fibers (see Methods). Figure 3A shows the measured EL intensity map of the heterostructure at the bias voltage of 2.0 V. Bright spots were exhibited at all seven indentation sites. The integrated intensities at the indentation sites were ~10 times higher than those at the surroundings without indents. A comparison of the EL before and after indentation reveals the following features. First, no point-like bright emission was observed before the indentation (fig. S2). This indicates that localized emission was induced only at the indentation sites. Second, the indentation changed the measured *I-V* curves (Fig. 3B). Whereas the turn-on voltage due to the tunneling current was ~4.0 V before the indentation, the indentation resulted in a lowering of the turn-on voltage to ~1.5 V and the appearance of an additional kink. These observations therefore show that the electrically driven excitons were efficiently funneled into individual exciton traps at the indentation sites, resulting in a decrease in the operating voltage.

We also measured the EL spectra at the indentation sites (Fig. 3C). The spectra exhibited the following features. First, narrow individual emission peaks with central wavelengths varying from 735 to 785 nm were observed. These wavelengths were longer than those of the charged or neutral excitons because the peaks originated from the localized exciton state. Second, the spectral linewidths of the peaks were almost two orders of magnitude narrower than those of the delocalized excitons. There were only one or two peaks with such sub-nanometer linewidths at each indentation site. Third, all seven indentation sites exhibited spectrally isolated peaks, while conventional defect-based TMDC light-emitting devices have one or two emitters over an area of ~40 $\mu m^2$ on average[24]. These characteristics of the emission peaks motivate the investigation of the emitter photon statistics.

To investigate the single-photon nature of the localized emissions from the indentation sites, we performed photon-correlation measurements using a Hanbury Brown and Twiss interferometer (see Methods). A narrow spectral window with the width of 10 nm was opened to select the desired peak in the EL spectra (black arrows in Fig. 3C). The spectrally filtered EL emission was split into two paths using a non-polarizing fiber beam splitter and detected by two APDs and time-tagging electronics that measured the time correlation between the detectors. We observed the single-photon properties from all seven indentation sites. Figure 3D shows three representative photon-correlation functions $g^{(2)}(\tau)$ measured at the indentation sites 1, 2, and 3 at the bias voltage of 2.0 V. The measured data were fitted with a three-level model curve (red lines in Fig. 3D). We then obtained the $g^{(2)}(0)$ values of $0.298 \pm 0.036$, $0.396 \pm 0.032$, and $0.353 \pm 0.041$ at the indentation sites 1, 2, and 3, respectively. These $g^{(2)}(0)$ values were below the threshold of 0.5 and exhibit the photon anti-bunching behavior expected in single-photon emission. The background EL emission was well suppressed, indicating a high purity of single-photon emission. We note that the measured $g^{(2)}$ data were not corrected for the background emission of the broad spectral window or the detector response that caused the non-zero value of $g^{(2)}(0)$; thus, the actual values for the purity of single-photon emission could be higher.

**Voltage-dependent EL measurements.** We next measured the EL spectrum at an indentation site (point 1 in Fig. 3A) under different bias voltages. The graphs in Fig. 4A show a strong emission peak at 739.8 nm with a narrow linewidth (<1 meV) at the bias voltages of 2.0 and 6.0 V. Only a single peak was observed at 2.0 V, whereas small side peaks appeared at 6.0 V. Other states, including unbound exciton emission, contributed to EL emission at a higher bias voltage[24-26]. In addition, we measured the evolution of the EL emission spectrum as the bias voltage increased (Fig. 4B). A single strong peak started to appear at ~1.5 V and its wavelength remained constant up to the bias voltage of 8 V. Additional peaks and broad background EL emission from the unbound exciton state appeared and became stronger as the bias voltage increased, but the strong peak at 739.8 nm was still dominant. This feature of the voltage-dependent EL spectrum was also reflected in the measured *I-V* characteristics shown in Fig. 3B. A strong single peak was observed at the turn-on voltage of >1.5 V, while the appearance of more emission peaks caused an additional kink at ~4 V in the *I-V* curve.

In addition, we plotted the measured peak intensity and linewidth from the indentation site 1 as a function of the bias voltage (Figs. 4C and D). As indicated in Fig. 2D, the EL emission started to emerge at a bias voltage of ~1.5 V and increased rapidly until ~2.5 V (Fig. 4C). Saturation behavior was also observed above ~2.5 V. The full-width-at-half-maximum (FWHM) of the EL emission increased from 0.164 nm (369 µeV) at ~1.5 V to 0.453 nm (936 µeV) at ~2.5 V as the bias voltage increased (Fig. 4D). Furthermore, the spectral stability of the sub-nanometer scale was measured for 5 min, which is required for practical single-photon emitters (fig. S3). Taken together, these measurement results indicate the single-photon nature of the electrically driven emission from the indentation sites.

**Fine-structure of single-photon emission.** A higher-resolution EL spectrum was next measured over the wavelength range of 738–742 nm to further investigate the optical properties of the dominant sharp emission peak (Fig. 5A). In the EL spectrum taken from indentation site 1 at 2.0 V, the fine-structure splitting was observed. There were two narrow peaks at the wavelengths of 739.5 and 740.0 nm with linewidths of ~400 µeV separated by 755 µeV. These narrow peaks were also observed at all seven indentation sites and showed similar emission characteristics, such as linewidths of a few hundred µeV and splittings of 700–1000 µeV. The zero-field splitting may possibly have originated from the anisotropic electron-hole exchange interaction caused by the low symmetry of the indentation sites[37-39]. In addition, we measured the polarization states of the two split peaks by placing a linear polarizer in front of the detector. The EL intensity was plotted as a function of the polarization direction (Fig. 5B). Strongly linear polarized states with the polarization degrees of 0.946 and 0.965 were observed at the lower (black) and higher (red) energy peaks, respectively. Notably, their polarization directions differed by 90°, and the photon energies were separated by ~800 μeV (Fig. 5C). A similar cross-polarized fine-structure splitting has also been observed in single neutral excitons localized in optically activated quantum emitters in TMDCs[1-3], as demonstrated by the energy level diagram with linearly polarized optical selection rules at zero magnetic field[6].

**Discussion**

In summary, we demonstrated electrically driven deterministic single-photon emitters in a vertical heterojunction with an active WSe$_2$ monolayer indented by an AFM tip. The employment of nanoscale elastic strain engineering to achieve exciton localization in the electronic structure enabled the efficient generation of robust single photons in our design. The fabrication process of the indentations was elucidated by numerical simulations. The turn-on voltage of the light-emitting heterostructure was manipulated through the band-structure engineering by local strain gradients. Indeed, we observed localized bright spots at the indentation sites, which exhibited narrow EL emission peaks with photon anti-bunching and intensity saturation behaviors at applied voltages of >1.5 V. We note that linear cross-polarized doublet in the localized emission can provide a valley degree of freedom as an information carrier.

Our fabrication procedure is scalable and applicable for integration with other photonic platforms such as high-quality optical cavities and/or low-loss optical waveguides. For example, the performance of single-photon emission can be further improved through Purcell enhancement by placing a nanoscale cavity on the indentation site. We believe that this work demonstrates an efficient approach for electrically driven single-photon sources in van der Waals heterostructure, outperforming the present emitters with random occurrence, and suggests next-generation single-photon devices for quantum information and quantum communication.

**Methods**

**Device fabrication.** Graphene/WSe$_2$/h-BN/graphene stacks on a polymer substrate were fabricated using typical mechanical exfoliation and transfer methods. Graphene monolayers grown on a copper film (2D Semiconductors) were first patterned using photolithography and oxygen plasma on a SiO$_2$/Si substrate. Mechanically exfoliated monolayer WSe$_2$ (HQ graphene) and few-layer h-BN were assembled on the other SiO$_2$/Si substrate, and monolayer graphene was placed on top via a dry transfer method. The stacked graphene/WSe$_2$/h-BN heterostructure was then placed on the other patterned graphene monolayer (fig. S1A). Next, PMMA was spin-coated on the heterostructure and the entire structure was peeled off using a water-assisted pick-up. The floated PMMA membrane was inverted and transferred onto another SiO$_2$/Si substrate, and annealed at 120 °C to make the PMMA layer uniform. Indentations were performed by adjusting the maximum cantilever displacement (i.e. z-piezo displacement) using commercial AFM equipment (Park Systems, NX10). We used the "Nanoindentation" mode in AFM to form nanoindents at the desired locations with the applied cantilever displacement of 1500 nm.

**Numerical simulation.** Mechanical contact simulations were performed using FEM (COMSOL Multiphysics, structural mechanics module) to calculate the deformation and strain field of the indented heterostructure on the elastic polymer substrate. The material properties of the polymer substrate were described by the Neo-Hookean hyperelastic material model: the density was 1,185 kg/m$^3$ and the Lamé parameter was 0.383 GPa. The indentation force was applied to the heterostructure using a contacted incompressible AFM tip with a radius of $r_t$. Based on the three-layer composite model in beam vibration theory, the TMDC heterostructure was assumed to be a single linear elastic material with the following effective mechanical parameters: the density was 2,950 kg/m$^3$, the Young's modulus was 884 GPa, and the Poisson's ratio was 0.2. The deformation of the indented heterostructure was then calculated as the force applied by the tip varied from 100 to 7,000 nN. The forces were determined based on the actual values used in the experiment. Furthermore, we obtained the displacement along the z-axis and calculated the absolute value of the diagonal strain tensor $|\varepsilon_{zz}|$.

**Optical measurements.** Optical measurements were performed on the samples using a home-built confocal microscope in a He-flow cryostat (Montana Instruments s50) at 4 K. For the high-resolution raster scans, two scanning galvo mirrors were used with a 4-f confocal alignment system. A bias voltage was applied to the samples using a source measurement unit (Keithley 2450). The EL emission was collected by a 100× objective lens with the numerical aperture of 0.85, and sent to either a monochromator/CCD (Princeton Instruments PIXIS 400 BRX) or avalanche photodiodes (Excelitas SPCM AQRH 13). Photon statistics measurement was performed using the HBT setup. The photons were spectrally filtered by band-pass filters, coupled to an optical fiber, split into two paths using a non-polarizing 50:50 fiber beam splitter, and finally detected by two identical avalanche photodiodes and time-tagging electronics (PicoQuant Picoharp 300). To perform polarization-resolved measurements, a linear polarizer was placed in front of the fiber coupler.

**Data availability.** The data that support the findings of this study are available from the corresponding author upon request.

**Acknowledgments**

This work was supported by the Samsung Research Funding & Incubation Center of Samsung Electronics (SRFC-MA2001-01).

**Author contributions**

J.-P.S. and H.-G.P. designed the experiments. J.-P.S., H.-R.K., H.-C.L., W.H., Y.S.K., K.W., T.T. and J.K. prepared the samples and performed experiments. J.-P.S. and H.-R.K. performed the simulations. H.B. and C.-H.L. analyzed the data. J.-P.S., H.B. and H.-G.P. wrote the paper. All authors discussed the results and commented on the manuscript.

**Competing interests**

The authors declare no competing interests.

**Correspondence and requests for materials** should be addressed to H.-G.P.

# References

1. A. Srivastava, M. Sidler, A. V. Allain, D. S. Lembke, A. Kis, A. Immamoglu, Optically active quantum dots in monolayer WSe$_2$. *Nat. Nanotechnol*. **10**, 491-496 (2015).

2. Y.-M. He, G. Clark, J. R. Schaibley, Y. He, M.-C. Chen, Y.-J. Wei, X. Ding, Q. Zhang, W. Yao, X. Xu, C.-Y. Lu, J.-W. Pan, Single quantum emitters in monolayer semiconductors. *Nat. Nanotechnol*. **10**, 497-502 (2015).

3. P. Tonndorf, R. Schmidt, R. Schneider, J. Kern, M. Buscema, G. A. Steele, A. Castellanos-Gomez, H. S. Zant, S. M. Vasconcellos, R. Bratschitsch, Single-photon emission from localized excitons in an atomically thin semiconductor. *Optica* **2**, 347-352 (2015).

4. D. White, A. Branny, R. J. Chapman, R. Picard, M. Brotons-Gisbert, A. Boes, A. Peruzzo, C. Bonato, B. D. Gerardot, Atomically-thin quantum dots integrated with lithium niobate photonic chips. *Opt. Mater. Express* **9**, 441-448 (2019).

5. T. Cai, J.-H. Kim, Z. Yang, S. Dutta, S. Aghaeimeibodi, E. Waks, Radiative enhancement of single quantum emitters in WSe$_2$ monolayers using site-controlled metallic nanopillars. *ACS Photon.* **5**, 3466-3471 (2018).

6. C. Chakraborty, N. Vamivakas, D. Englund, Advances in quantum light emission from 2D materials. *Nanophotonics* **8**, 2017-2032 (2019).

7. C. Palacios-Berraquero, D. M. Kara, A. R. Montblanch, M. Barbone, P. Latawiec, D. Yoon, A. K. Ott, M. Loncar, A. C. Ferrari, M. Atatüre, Large-scale quantum-emitter arrays in atomically thin semiconductors. *Nat. Commun*. **8**, 15093 (2017).

8. A. Branny, S. Kumar, R. Proux, B. D. Gerardot, Deterministic strain-induced arrays of quantum emitters in a two-dimensional semiconductor. *Nat. Commun*. **8**, 15053 (2017).

9. G. D. Shepard, O. A. Ajayi, X. Li, X.-Y. Zhu, J. Hone, S. Strauf, Nanobubble induced formation of quantum emitters in monolayer semiconductors. *2D Mater*. **4**, 021019 (2017).

10. F. Peyskens, C. Chakraborty, M. Muneeb, D. V. Thourhout, D. Englund, Integration of single photon emitters in 2D layered materials with a silicon nitride photonic chip. *Nat. Commun*. **10**, 4435 (2019).

11. M. Blauth, M. Jürgensen, G. Vest, O. Hartwig, M. Prechtl, J. Cerne, J. J. Finley, M Kaniber, Coupling single photons from discrete quantum emitters in WSe$_2$ to lithographically defined plasmonic slot waveguides. *Nano Lett*. **18**, 6812-6819 (2018).

12. T. Cai, S. Dutta, S. Aghaeimeibodi, Z. Yang, S. Nah, J. T. Fourkas, E. Waks, Coupling

emission from single localized defects in two-dimensional semiconductor to surface plasmon polaritons. *Nano Lett.* **17**, 6564-6568 (2017).

13. P. Tonndorf, O. D. Pozo-Zamudio, N. Gruhler, J. Kern, R. Schmidt, A. I. Dmitriev, A. P. Bakhtinov, A. I. Tartakovskii, W. Pernice, S. M. Vasconcellos, R. Bratschistsch, On-chip waveguide coupling of a layered semiconductor single-photon source. *Nano Lett.* **17**, 5446-5451 (2017).
14. J. Kern, I. Niehues, P. Tonndorf, R. Schmidt, D. Wigger, R. Schneider, T. Stiehm, S. M. Vasconcellos, D. E. Reiter, T. Kuhn, R. Bratschitsch, Nanoscale positioning of single-photon emitters in atomically thin $WSe_2$. *Adv. Mater*. **28**, 7101-7105 (2016).
15. Y. Luo, G. D. Shepard, J. V. Ardelean, D. A. Rhodes, B. Kim, K. Barmak, J. C. Hone, S. Strauf, Deterministic coupling of site-controlled quantum emitters in monolayer $WSe_2$ to plasmonic nanocavities. *Nat. Nanotechnol.* **13**, 1137-1142 (2018).
16. L. Linhart, M. Paur, V. Smejkal, J. Burgdörfer, T. Mueller, F. Libisch, Localized intervalley defect excitons as single-photon emitters in $WSe_2$. *Phys. Rev. Lett.* **123**, 146401 (2019).
17. J.-P. So, K.-Y. Jeong, J. M. Lee, K.-H. Kim, S.-J. Lee, W. Huh, H.-R. Kim, J.-H. Choi, J. M. Kim, Y. S. Kim, C.-H. Lee, S. Nam, H.-G. Park, Polarization control of deterministic single-photon emitters in monolayer $WSe_2$. *Nano Lett*. **21**, 1546-1554 (2021).
18. A. Hötger, J. Klein, K. Barthelmi, L. Sigl, F. Sigger, W. Männer, S. Gyger, M. Florian, M. Lorke, F. Jahnke, T. Taniguchi, K. Watanabe, K. D. Jöns, U. Wurstbauer, C. Kastl, K. Müller, J. J. Finley, A. W. Holleitner, Gate-switchable arrays of quantum light emitters in contacted monolayer $MoS_2$ van der Waals heterodevices. *Nano Lett*. **21**, 1040-1046 (2021).
19. C. Chakraborty, K. M. Goodfellow, S. Dhara, A. Yoshimura, V. Meunier, A. N. Vamivakas, Quantum-confined Stark effect of individual defects in a van der Waals heterostructure. *Nano Lett*. **17**, 2253-2258 (2017).
20. C. Chakraborty, L. Kinnischtzke, K. M. Goodfellow, R. Beams, A. N. Vamivakas, Voltage-controlled quantum light from an atomically thin semiconductor. *Nat. Nanotechnol.* **10**, 507-511 (2015).
21. A. K. Geim, I. V. Grigorieva, Van der Waals heterostructures. *Nature* **499**, 419-425 (2013).
22. F. Withers, O. D. Pozo-Zamudio, A. Mishchenko, A. P. Rooney, A. Gholinia, K. Watanabe, T. Taniguchi, S. J. Haigh, A. K. Geim, A. I. Tartakovskii, K. S. Novoselov, Light-emitting diodes by band-structure engineering in van der Waals heterostructures. *Nat. Mater.* **14**, 301-

306 (2015).

23. F. Withers, O. D. Pozo-Zamudio, S. Schwarz, S. Dufferwiel, P. M. Walker, T. Godde, A. P. Rooney, A. Gholinia, C. R. Woods, P. Blake, S. J. Haigh, K. Watanabe, T. Taniguchi, I. L. Aleiner, A. K. Geim, V. I. Fal'ko, A. I. Tartakovskii, K. S. Novoselov, $WSe_2$ light-emitting tunneling transistors with enhanced brightness at room temperature. *Nano Lett*. **15**, 8223-8228 (2015).

24. C. Palacios-Berraquero, M. Barbone, D. M. Kara, X. Chen, I. Goykhman, D. Yoon, A. K. Ott, J. Beitner, K. Watanabe, T. Taniguchi, A. C. Ferrari, M. Atatüre, Atomically thin quantum light-emitting diodes. *Nat. Commun.* **7**, 12978 (2016).

25. G. Clark, J. R. Schaibley, J. Ross, T. Taniguchi, K. Watanabe, J. R. Hendrickson, S. Mou, W. Yao, X. Xu, Single defect light-emitting diode in a van der Waals heterostructure. *Nano Lett*. **16**, 3944-3948 (2016).

26. S. Schwarz, A. Kozikov, F. Withers, J. K. Maguire, A. P. Foster, S. Dufferwiel, L. Hague, M. N. Makhonin, L. R. Wilson, A. K. Geim, K. S. Novoselov, A. I. Tartakovskii, Electrically pumped single-defect light emitters in $WSe_2$. *2D Mater*. **3**, 025038 (2016).

27. M. R. Rosenberger, C. K. Dass, H.-J. Chuang, S. V. Sivaram, K. M. McCreary, J. R. Hendrickson, B. T. Jonker, Quantum calligraphy: Writing single-photon emitters in a two-dimensional materials platform. *ACS Nano* **13**, 904-912 (2019).

28. B. W. H. Baugher, H. O. H. Churchill, Y. Yang, P. Jarillo-Herrero, Optoelectronic devices based on electrically tunable p-n diodes in a monolayer dichalcogenide. *Nat. Nanotechnol*. **9**, 262-267 (2014).

29. A. Pospischil, M. M. Furchi, T. Mueller, Solar-energy conversion and light emission in an atomic monolayer p-n diode. *Nat. Nanotechnol.* **9**, 257-261 (2014).

30. J. S. Ross, P. Klement, A. M. Jones, N. J. Ghimire, J. Yan, D. G. Mandrus, T. Taniguchi, K. Watanabe, K. Kitamura, W. Yao, D. H. Cobden, X. Xu, Electrically tunable excitonic light-emitting diodes based on monolayer $WSe_2$ p-n junctions. *Nat. Nanotechnol*. **9**, 268-272 (2014).

31. D.-H. Lien, M. Amani, S. B. Desai, G. H. Ahn, K. Han, J.-H. He, J. W. Ager III, M. C. Wu, A. Javey, Large-area and bright pulsed electroluminescence in monolayer semiconductors. *Nat. Commun*. **9**, 1229 (2018).

32. P. Kumar, V. Balakrishnan, Nanosculpting of atomically thin 2D materials for site-specific photoluminescence modulation. *Adv. Optical. Mater*. **6**, 1701284 (2018).

33. A. Castellanos-Gomez, M. Buscema, R. Molenaar, V. Singh, L. Janssen, H. S. Zant, G. A. Steele, Deterministic transfer of two-dimensional materials by all-dry viscoelastic stamping. *2D Mater.* **1**, 011002 (2014).

34. K.-H. Cho, Y. Kim, Elastic modulus measurement of multilayer metallic thin films. *J. Mater. Res.* **14**, 1996-2001 (1999).

35. D. Akinwande, N. Petrone, J. Hone, Two-dimensional flexible nanoelectronics. *Nat. Commun*. **5**, 5678 (2014).

36. R. C. Cooper, C. Lee, C. A. Marianetti, X. Wei, J. Hone, J. W. Kysar, Nonlinear elastic behavior of two-dimensional molybdenum disulfide. *Phys. Rev. B*. **87**, 035423 (2013).

37. C. Chakraborty, N. R. Jungwirth, G. D. Fuchs, A. N. Vamivakas, Electrical manipulation of the fine-structure splitting of $WSe_2$ quantum emitters. *Phys. Rev. B*. **99**, 045308 (2019).

38. C, Chakraborty, A. Mukherjee, L. Qiu, A. N. Vamivakas, Electrically tunable valley polarization and valley coherence in monolayer $WSe_2$ embedded in a van der Waals heterostructure. *Opt. Mater. Express.* **9**, 1479-1487 (2019).

39. S. Kumar, M. Brotons-Gisbert, R. Al-Khuzheyri, A. Branny, G. Ballesteros-Garcia, J. F. Sanchez-Royo, B. D. Gerardot, Resonant laser spectroscopy of localized excitons in monolayer $WSe_2$. *Optica* **3**, 882-886 (2016).

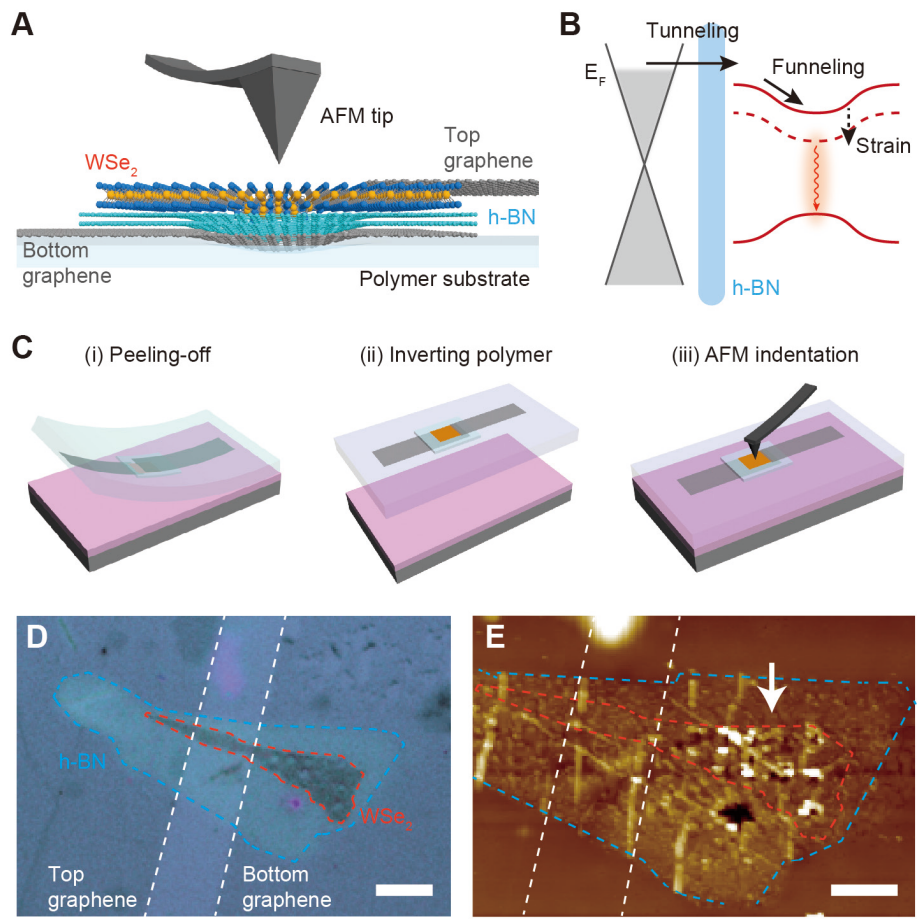

**Fig. 1. Electrically driven single-photon emitter formed by AFM indentation.** (**A**) Schematic of the vertical heterostructure consisting of top-graphene/WSe$_2$/h-BN/bottom-graphene on the polymer substrate. The emission site is indented by an AFM tip. (**B**) Schematic band diagram of the strained heterostructure under external bias. The Fermi level ($E_F$) of the monolayer bottom graphene (gray) is raised above the conduction band edge of the monolayer WSe$_2$. This allows electrons to tunnel from the bottom graphene to WSe$_2$ through the h-BN. The exciton trap in the bandgap of WSe$_2$ is modulated by the local strain (red), resulting in single-photon emission. (**C**) Schematics of the fabrication procedures: (i) The PMMA layer was spin-coated on top of the TMDC heterostructure on the SiO$_2$/Si substrate and peeled off together with the heterostructure from the substrate. (ii) The detached PMMA layer was turned upside down while floating on DI

water, and placed and aligned on another substrate. (iii) AFM indentation was performed to apply a strain to the heterostructure on the PMMA layer. (**D**) Optical microscope image of the heterostructure placed on the PMMA substrate. The dashed lines highlight the footprints of the monolayer top graphene and bottom graphene (white), the few layer h-BN (blue), and the monolayer $WSe_2$ (red). Scale bar, 10 μm. (**E**) AFM image of the fabricated heterostructure. The indents were formed on the stacked $WSe_2$ area (white arrow). Scale bar, 5 μm.

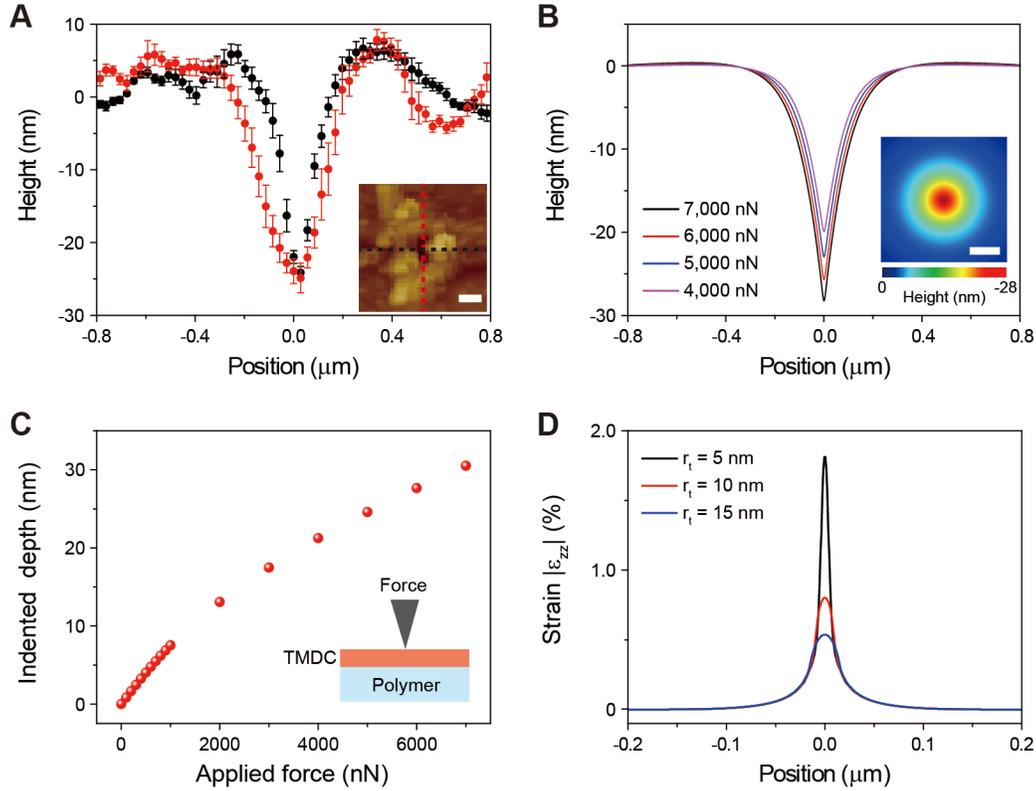

**Fig. 2. Mechanical properties of the indented heterostructure.** (**A**) Measured line profiles of the heights averaged from seven indents formed along the lines (black and red) labeled in the inset image. The error bars indicate the standard deviation. The inset shows a magnified image of the indents in Fig. 1E. Scale bar, 200 nm. (**B**) Simulated morphology of the heterostructure indented at various applied forces using a tip with the radius of 5 nm. The inset image shows the top-view of the height profile for the applied force of 7,000 nN. Scale bar, 100 nm. (**C**) Simulated maximum indented depth as a function of the applied force. The inset shows the schematic of the simulation where the TMDC heterostructure is modeled as a single linear elastic material with effective mechanical parameters. (**D**) Calculated diagonal component of the strain along the vertical direction, $|\varepsilon_{zz}|$, for the three different tip radii of 5, 10, and 15 nm. Continuum theory was used for an elastic plate as follows: $|\varepsilon_{zz}| = \left|\frac{vt}{1-v}(\frac{\partial^2 h}{\partial x^2} + \frac{\partial^2 h}{\partial y^2})\right|$, where $h$ is the height profile of WSe$_2$, $t$ is the thickness of WSe$_2$, and $v$ is Poisson's ratio. We set $t = 0.7$ nm and $v = 0.19$.

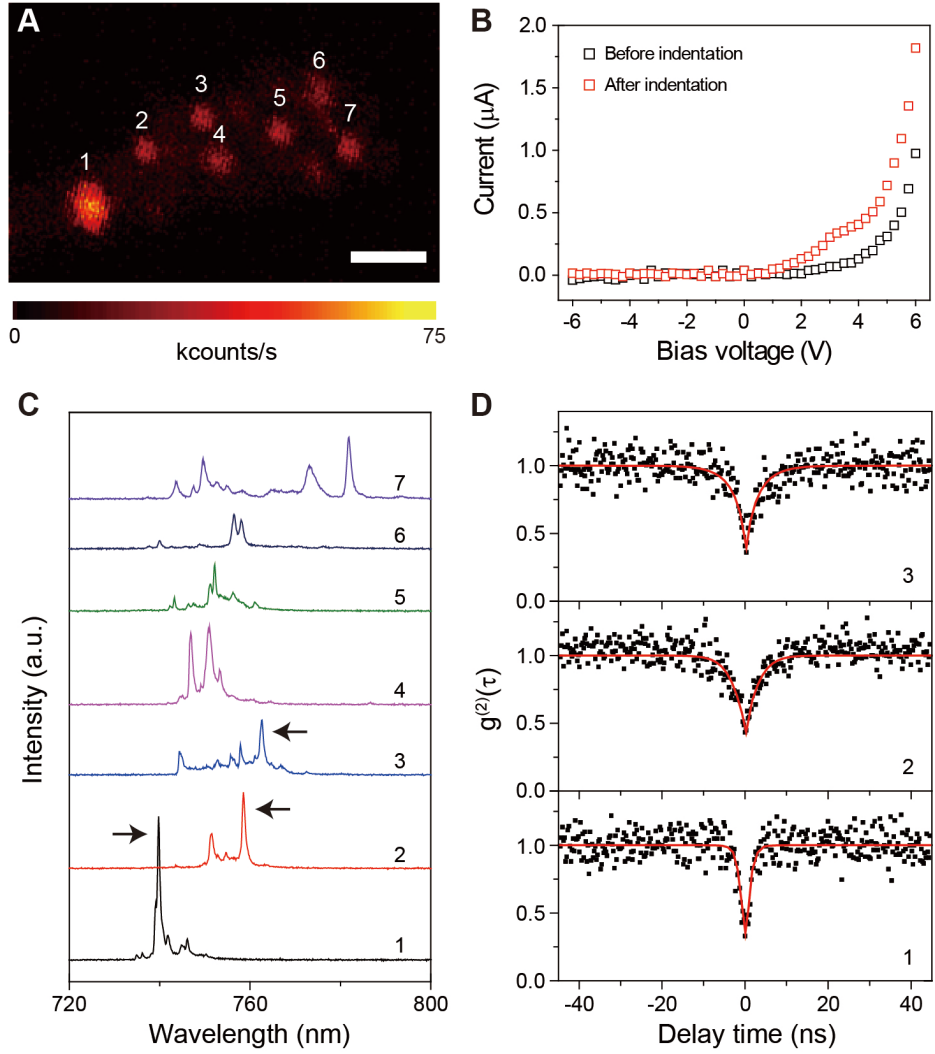

**Fig. 3. Measured optoelectronic properties at the indentation sites.** (**A**) EL intensity map measured in the heterostructure in Fig. 1E. Localized emissions were observed at indentation sites 1 to 7. The bias voltage was 2.0 V. Scale bar, 2 μm. (**B**) Plot of current vs. voltage at 4 K before (black) and after (red) indentation. (**C**) EL spectra at the indentation sites labeled in (A). (**D**) Measured second-order correlation function $g^{(2)}(\tau)$ at the indentation sites 1, 2, and 3 in (C). The bias voltage was 2.0 V, and a 10 nm-wide spectral filter was used. The fitted red curves indicate photon anti-bunching and correspond to the $g^{(2)}(0)$ values of $0.298 \pm 0.036$ (bottom), $0.396 \pm 0.032$ (middle), and $0.353 \pm 0.041$ (top), respectively.

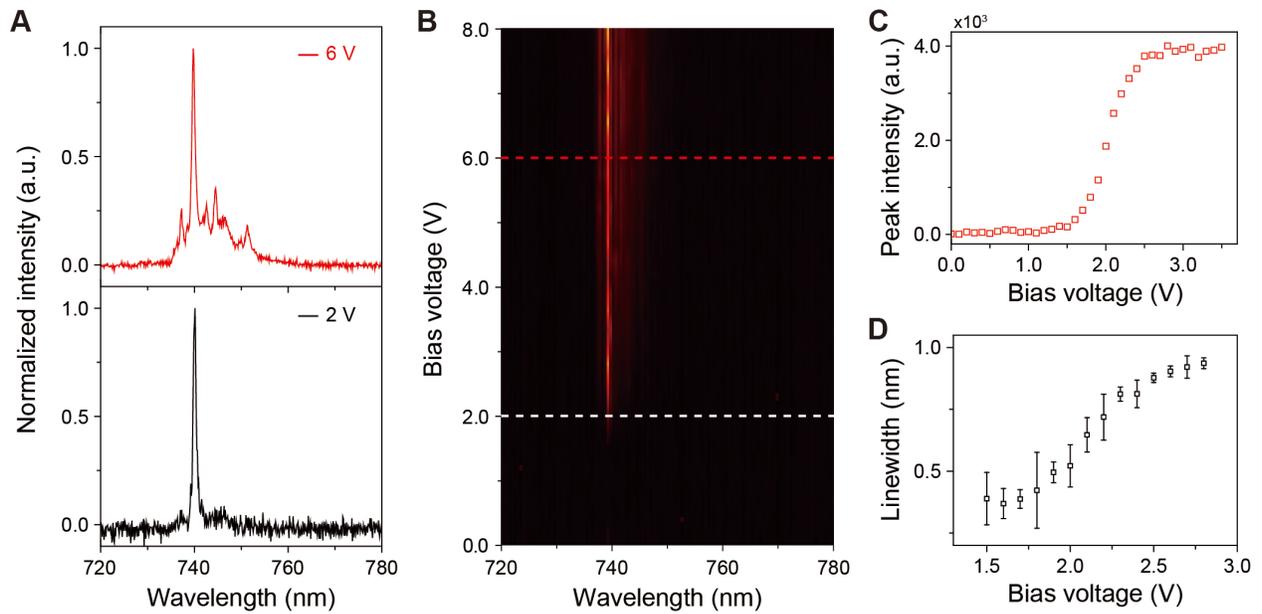

**Fig. 4. EL emission at different bias voltages.** (**A**) Normalized EL spectra measured at indentation site 1 at the bias voltages of 2.0 V (bottom) and 6.0 V (top). (**B**) Plot of bias voltage vs. wavelength of the EL spectrum. (**C**) Measured EL intensity as a function of the bias voltage. Intensity saturation was observed at biases above 2.5 V. (**D**) Measured peak linewidth as a function of the bias voltage at indentation site 1 in Fig. 3A. The error bars indicate the standard deviation.

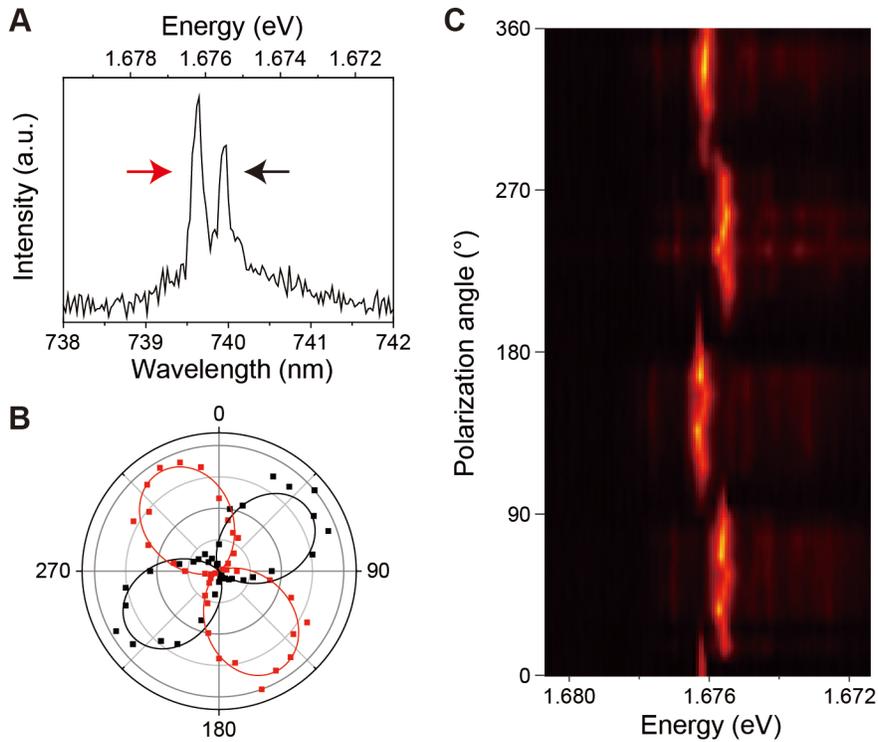

**Fig. 5. Fine-structure splitting of the single-photon emission.** (**A**) High-resolution EL spectrum measured at indentation site 1 labeled in Fig. 3A. The fine-structure splitting has a separation of 755 μeV and a linewidth of ~400 μeV. (**B**) Polar plots of the normalized EL intensities of the lower (black) and higher (red) energy peaks, which are denoted by arrows with the corresponding colors in (A), as a function of the polarization angle. The lines indicate the fitted curves. The polarization angles and degrees are 56° and 0.946 (black) and 148° and 0.965 (red), respectively. (**C**) EL spectrum of (A) as a function of the polarization angle.

*Supplementary Information for:*

# Electrically driven strain-induced deterministic single-photon emitters in a van der Waals heterostructure

Jae-Pil So[1], Ha-Reem Kim[1], Hyeonjun Baek[1,2], Hoo-Cheol Lee[1], Woong Huh[3], Yoon Seok Kim[3], Kenji Watanabe[4], Takashi Taniguchi[5], Jungkil Kim[6], Chul-Ho Lee[3], and Hong-Gyu Park[1,3*]

*Corresponding author. E-mail: hgpark@korea.ac.kr

This PDF file includes:

Supplementary Figures S1–S3

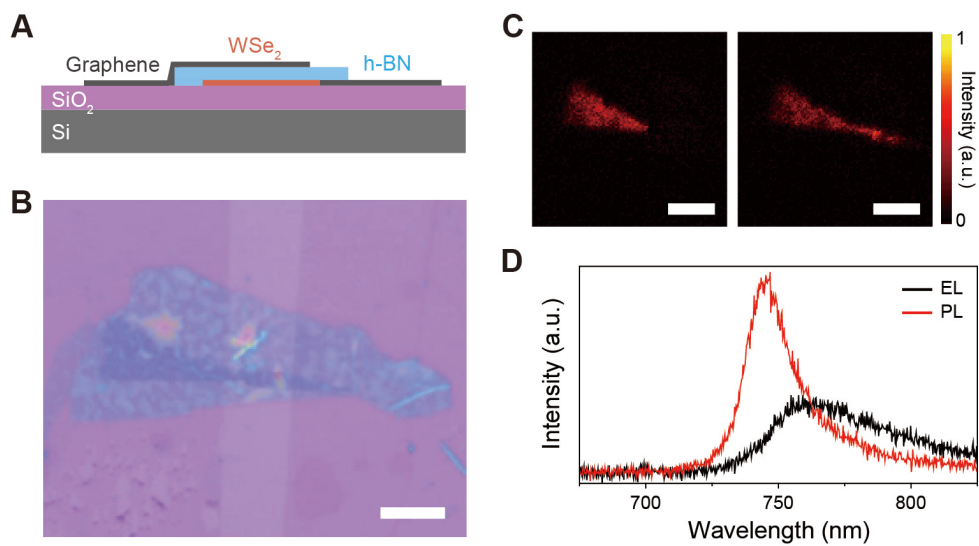

**Fig. S1. 2D heterostructure on SiO$_2$/Si substrate.** (**A**) Schematic of the heterostructure. (**B**) Optical microscope image of the fabricated structure. Scale bar, 5 μm. (**C**) EL (left) and PL (right) intensity maps measured at room temperature for the structure in (B). Scale bar, 5 μm. (**D**) EL (black) and PL (red) spectra for the structure in (B).

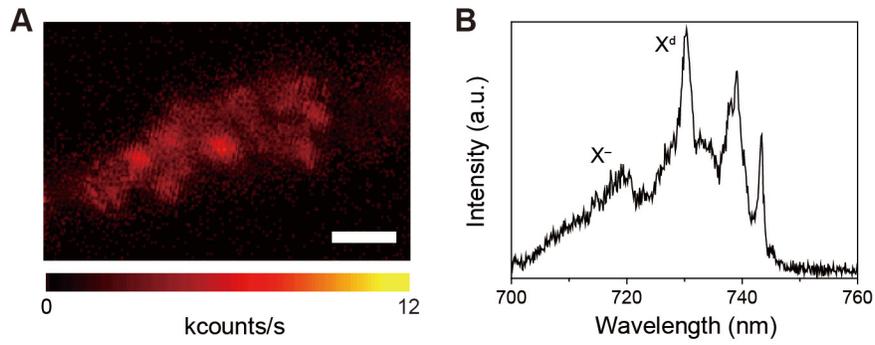

**Fig. S2. Measurement of EL emission in the heterostructure without indentation sites.** (**A**) Measured EL intensity map at 4 K for the structure in Fig. 1D before indentation. The emission was delocalized and roughly uniform. The bias voltage was 8.0 V. Scale bar, 2 μm. (**B**) Measured EL spectrum in (A). The peaks for charged excitons (marked as $X^-$) and weakly defect-bound excitons (marked as $X^d$) are shown.

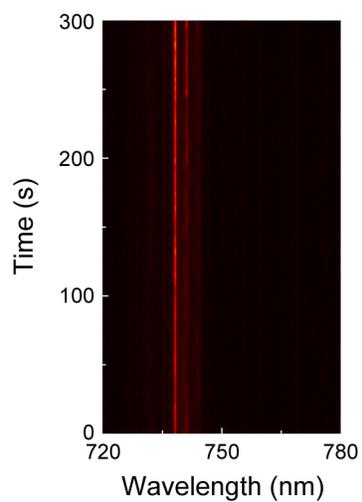

**Fig. S3. Spectral stability.** The EL spectrum was measured as a function of time over a duration of 5 min. The dominant narrow peak occurred at ~740 nm. We acquired the spectrum data every 1 s. This result confirms the spectral stability of the single-photon emission at the indentation site.